# Prediction of multivariate responses with a select number of principal components

**Inge Koch**

*School of Mathematics and Statistics*
*University of New South Wales*
*Australia*
*e-mail:* `inge@unsw.edu.au`

and

**Kanta Naito**

*Department of Mathematics*
*Shimane University*
*Japan*
*e-mail:* `naito@riko.shimane-u.ac.jp`

**Abstract:** This paper proposes a new method and algorithm for predicting multivariate responses in a regression setting. Research into classification of High Dimension Low Sample Size (HDLSS) data, in particular microarray data, has made considerable advances, but regression prediction for high-dimensional data with continuous responses has had less attention. Recently Bair *et al* (2006) proposed an efficient prediction method based on supervised principal component regression (PCR). Motivated by the fact that a larger number of principal components results in better regression performance, this paper extends the method of Bair *et al* in several ways: a comprehensive variable ranking is combined with a selection of the best number of components for PCR, and the new method further extends to regression with multivariate responses. The new method is particularly suited to HDLSS problems. Applications to simulated and real data demonstrate the performance of the new method. Comparisons with Bair *et al* (2006) show that for high-dimensional data in particular the new ranking results in a smaller number of predictors and smaller errors.

## 1. Introduction

Classification of high-dimensional data – motivated mainly by the importance of tumor classifications for high-dimensional microarray data – has attracted much recent attention. To date less attention has been paid to the prediction of survival times for gene expression data, although such prediction can provide valuable additional knowledge and important information in the selection of relevant genes. One important advance in this area is Bair *et al* (2006) who proposed a prediction method for multiple regression based on supervised principal components. They proposed not only a new method, but also an underlying model based on latent variables which provides a good and theoretically founded explanation of their method.

In this paper we focus on prediction of continuous response variables in a very general linear regression framework and extend the research of Bair *et al* (2006) in a number of crucial ways. Like Bair *et al* we allow a very large number of predictors, but in addition we consider multivariate reponses as met in practice, for example, in the prediction of treatment placement and facilities in drug related offences. Including multivariate responses in our model is a genuine extension over Bair *et al* whose approach cannot handle such data.

A central problem in linear regression from multivariate predictors is the choice of variables that are included in the prediction model. Section 2 briefly reviews the classical case and principal component regression (PCR). For high-dimensional data PCR has an inuitive appeal for regression prediction, and we follow that path - similar to Bair *et al* (2006). But unlike their approach, which uses the first principal component only, we integrate the dimension selection of Koch and Naito (2007) into PCR, which yields a better fit based on more than one principal component.

Compared to Bair *et al*, our prediction model is more complex and extensive; our choice of dimension has a theoretical foundation, and since we use more than one predictor, our prediction model results in more accurate prediction.

Another prediction method with applications in chemometrics has been proposed in Gustafsson (2005). This approach also falls into the PCR framework but then uses a further rotation of the sphered principal components. For regression predictions such rotations have no effect, since they would cancel out in an appropriate verson of (5). For this reason we compare our results to those of Bair *et al* (2006).

A key issue in classification and prediction of a continuous response variable from high-dimensional data is the preselection of a moderate number of variables that have strong predictive power. In microarray analysis this first selection is often achieved by univariate t-tests. Bair *et al* (2006) replace these tests, essentially by calculating the univariate correlation coefficient between each predictor and the response variable, and make a preselection of the variables based on the absolute value of these correlation coefficients. This preselection is simple to implement and understand, but does not capture the interaction of the variables or genes. We address the preselection problem and propose a ranking of the variables which takes into account all predictors simultaneously, and thus takes care of valuable interaction between the variables in the preselection.

Our prediction method can be regarded as a two-step variable selection. The first step selects the variables that are most important for prediction, while the second step summarises these variables in a smaller, and judiciously chosen, number of components.

The paper is organised as follows. Section 2 reviews regression prediction, and describes variable ranking based on a canonical correlation approach. Section 3 proposes our regression algorithm which is based on variable ranking and dimension selection. Section 4 provides more details and properties of our variable selection, and includes comparisons with the method of Bair *et al* (2006). Section 5 shows how our method performs in practice for simulated and real data, including an example of high-dimensional microarray data, and gives comparisons with the approach of Bair *et al* (2006). We conclude with a discussion of our method and algorithm in Section 6.

## 2. Prediction and Canonical Correlations

### 2.1. Regression Prediction

We consider the multivariate regression setting:

$$\mathbf{Y} = \mathbf{X}\mathbf{B} + \mathbf{E}, \tag{1}$$

where $\mathbf{Y} = [\mathbf{y}_1 \cdots \mathbf{y}_N]^T$ denotes the matrix of responses, and, each response $\mathbf{y}_i$ is a vector of length $q$. Further, $\mathbf{B} = [\boldsymbol{\beta}_1 \cdots \boldsymbol{\beta}_p]^T$ is the $p \times q$ matrix of coefficients, and $\mathbf{E}$ is the $N \times q$ error matrix. The design matrix is the same as in a multiple regression setting: $\mathbf{X} = [\mathbf{x}_1 \cdots \mathbf{x}_p]$, so each variable or feature vector $\mathbf{x}_j$ is of length $N$. Throughout this paper we will assume that the columns of $\mathbf{X}$ are centred. An estimator for $\mathbf{B}$ is given by

$$\widehat{\mathbf{B}} = (\tilde{\mathbf{X}}^T \tilde{\mathbf{X}})^{-1} \tilde{\mathbf{X}}^T \mathbf{Y}, \tag{2}$$

where $\tilde{\mathbf{X}}$ is derived from $\mathbf{X}$, and a new $q$-dimensional response $\widehat{\mathbf{y}}$ is predicted for a single $p$-dimensional datum $\mathbf{x}_0$ by

$$\widehat{\mathbf{y}}^T = \mathbf{x}_0^T (\tilde{\mathbf{X}}^T \tilde{\mathbf{X}})^{-1} \tilde{\mathbf{X}}^T \mathbf{Y}. \tag{3}$$

If $p < N$ and $(\mathbf{X}^T\mathbf{X})^{-1}$ exists, then $\tilde{\mathbf{X}} = \mathbf{X}$ leads to an estimator and new predictions based on all variables. If the inverse $(\mathbf{X}^T\mathbf{X})^{-1}$ does not exist or is unstable, then a smoothed or penalised inverse as in ridge regression or the more recent lasso - see Hastie *et al* (2007) or Meier and Bühlmann (2007) and references therein - is commonly used. In this paper we shall not be concerned with such inverses.

In multiple linear regression the aim is to reduce the number of variables. Without going into the different methods of finding such subsets of variables, but assuming instead one has found a subset $\mathbf{X}^-$ where the superscript $^-$ indicates that some variables are omitted, then $\tilde{\mathbf{X}} = \mathbf{X}^-$ and $\tilde{\mathbf{x}}_0 = \mathbf{x}_0^-$ in (2) and (3).

PCR does not leave out variables, but uses weighted sums of all variables, with weights allocated according to the contribution to variance of the variables. Let

$$\mathbf{Z}^{(k)} = \mathbf{X}\boldsymbol{\Gamma}_k = [\mathbf{z}_1 \cdots \mathbf{z}_k]. \tag{4}$$

Then $\mathbf{Z}^{(k)}$ is an $N \times k$ matrix, with $k \leq p$, $\boldsymbol{\Gamma}_k$ denotes the matrix which consists of the first $k$ eigenvectors of the sample covariance matrix of the predictors, and the $i$-th column vector $\mathbf{z}_i$ is obtained by projecting $\mathbf{X}$ onto the $i$-th eigenvector $\gamma_i$ of $\Gamma_k$. In the PCR setting $\tilde{\mathbf{X}} = \mathbf{Z}^{(k)}$ and $\tilde{\mathbf{x}}_0 = \mathbf{z}_0^{(k)}$, and one obtains the estimator $\widehat{\mathbf{B}}$ and predictor $\widehat{\mathbf{y}}$ given by

$$\begin{aligned}
\widehat{\mathbf{B}} &= \left(\mathbf{Z}^{(k)T}\mathbf{Z}^{(k)}\right)^{-1}\mathbf{Z}^{(k)T}\mathbf{Y} = \left(\Gamma_k^T \mathbf{X}^T \mathbf{X} \Gamma_k\right)^{-1}\Gamma_k^T \mathbf{X}^T \mathbf{Y} \quad \text{and} \\
\widehat{\mathbf{y}}^T &= \mathbf{x}_0^T \Gamma_k \left(\Gamma_k^T \mathbf{X}^T \mathbf{X} \Gamma_k\right)^{-1}\Gamma_k^T \mathbf{X}^T \mathbf{Y}.
\end{aligned} \quad (5)$$

Since $\mathbf{z}_1$ is a linear combination of all variables, $k=1$ is often used in PCR, for example in Bair *et al* (2006). This first principal component often does not sufficiently summarise the predictors, and then it is important to find an appropriate value for $k$. We return to this choice in Section 3.5.2.

### 2.2. Variable Ranking and Canonical Correlations

In univariate regression the correlation coefficient $\rho$ of $X$ and $Y$ is defined by

$$\rho = \frac{cov(X,Y)}{\sqrt{var[X]var[Y]}}, \quad (6)$$

where *cov* refers to the covariance and *var* to the variance. For regression, this coefficient relates the standardised response and predictor variables.

In multiple linear regression the absolute value of the correlation coefficient can be used to order the predictor variables according to the strength of their correlation with a univariate response $Y$, so

$$[\mathbf{x}_1 \cdots \mathbf{x}_p] \quad \to \quad [\mathbf{x}_{(1)} \cdots \mathbf{x}_{(p)}]$$

where the vectors $\mathbf{x}_{(i)}$ are ordered such that $|\rho_{(1)}| \geq |\rho_{(2)}| \cdots |\rho_{(p)}|$, and $\rho_{(i)}$ denotes the correlation between a response vector $\mathbf{y}$ and $\mathbf{x}_{(i)}$. As pointed out in Section 1, the correlation coefficient takes into account one feature vector at a time - so considers only the marginals - and cannot account for any interaction between the $\mathbf{x}_j$s. Further, the correlation coefficients and the above ordering do not extend to multivariate responses.

For a multivariate predictor $\boldsymbol{X}$ and a multivariate response $\boldsymbol{Y}$ a natural generalisation of $\rho$ is the matrix of canonical correlations

$$C = \Sigma_X^{-1/2} \Sigma_{XY} \Sigma_Y^{-1/2}, \quad (7)$$

which replaces the univariate $cov(X,Y)$ of (6) by the $p \times q$ covariance matrix $\Sigma_{XY}$ of $\boldsymbol{X}$ and $\boldsymbol{Y}$, and each of $var[X]$ and $var[Y]$ by their respective variance matrices $\Sigma_X$ and $\Sigma_Y$. The matrix $C$ is commonly used in canonical correlation analysis, and contains information about the strength of the relationship of the individual variables. For a univariate response $Y$, the matrix $C$ reduces to a $p \times 1$ vector whose entries give rise to an ordering or ranking of the $p$ variables of $\mathbf{X}$ which makes use of all interactions of the data. In Step 1 of our algorithm we show how the matrix $C$ induces a a ranking of the data variables which also applies

1. to multivariate responses **Y**, and
2. to data whose dimension $p$ exceeds the sample size $N$.

## 3. A Regression Algorithm Based on Ranking and Dimension Selection

### 3.1. The Underlying Latent Variable Model

The model for a single response vector **y** is

$$\mathbf{y} = \mathbf{W}^T \mathbf{s} + \boldsymbol{\delta}, \tag{8}$$

where $\mathbf{W}^T$ is a $q \times H$ matrix, **s** is an $H \times 1$ latent vector with non-Gaussian components, $H \ll p$, and $\boldsymbol{\delta}$ is a $q \times 1$ error vector. Let $I$ denote a proper subset of $\{1,...,p\}$ and set $|I| = m < p$. For $i \in I$, the $i$-th feature of the input vector **x** is modelled by

$$x_i = \mathbf{p}_i^T \mathbf{s} + \eta_i, \tag{9}$$

where the $\mathbf{p}_i$s are $H$-dimensional vectors, the $\eta_i$ are errors, and $H \leq m$. Combining (8) and (9), we have

$$\begin{bmatrix} \mathbf{x}_* \\ \mathbf{y} \end{bmatrix} = \begin{bmatrix} \mathbf{P} \\ \mathbf{W}^T \end{bmatrix} \mathbf{s} + \begin{bmatrix} \boldsymbol{\eta} \\ \boldsymbol{\delta} \end{bmatrix}, \tag{10}$$

where $\mathbf{x}_*$ is a $m \times 1$ subvector of **x** defined by the components of (9) and **P** is a $m \times H$ matrix having $\mathbf{p}_i^T (i \in I)$ as its rows. The $m$-dimensional vector $\mathbf{x}_*$ contains the features which are most relevant for prediction and which are modelled by a smaller number $H$ of latent variables. Step 1 of the algorithm shows how we obtain the vector $m$, and Step 2 elucidates the choice of $H$.

### 3.2. Variable Ranking with C

We begin with an inspection of the matrix $C$ given in (7) which can be regarded as a multivariate correlation coefficient of size $p \times q$. To be able to rank the variables of **X**, we use the fact that the strongest correlation between the **X** and **Y** variables is given by the largest eigenvalue, $\kappa_1$, of $C$.

The first left and right eigenvectors of $C$, denoted by $\mathbf{h}_1$ and $\mathbf{g}_1$ respectively, satisfy

$$C\mathbf{g}_1 = \kappa_1 \mathbf{h}_1. \tag{11}$$

Further the entries of $\mathbf{h}_1$ and $\mathbf{g}_1$ contain relative weights for the variables of **X** and **Y** (see Theorem 3.6, Chapter 3 of Koch (2009)). Next observe that the correlation coefficient $\rho$ relates the univariate standardised variables $X$ and $Y$ via $Y/\sigma_Y = \rho X/\sigma_X$, where $\sigma_Y$ and $\sigma_X$ is the standard deviation of $Y$ and $X$, respectively, and both $X$ and $Y$ are assumed to be centred. The multivariate analogue of this relationship is

$$\begin{aligned} \Sigma_Y^{-1/2} \mathbf{Y} &= C^T \Sigma_X^{-1/2} \mathbf{X} \quad \text{or equivalently} \\ \mathbf{Y} &= \tilde{C}^T \mathbf{X}, \end{aligned} \tag{12}$$

where $\tilde{C} = \Sigma_X^{-1/2} C \Sigma_Y^{1/2} = \Sigma_X^{-1} \Sigma_{XY}$. Combining (11) and (12) we note that $C$ applies to the sphered data while $\tilde{C}$ applies directly to the raw data. Put

$$\mathbf{b} = \frac{1}{\kappa_1} \Sigma_X^{-1/2} C \mathbf{g}_1, \tag{13}$$

and observe that $\mathbf{b}$ is the first (left) canonical covariate vector (see Chapter 3, Koch (2009)).

Ranking of variables is particularly important for HDLSS data, when the number of variables exceeds the number of observations. In this case the usual estimate of the covariance matrix $\Sigma_X$ is not invertible. Assume that $\mathbf{X}$ has rank $r$ (with $r \leq \min(N, p)$). Let

$$\mathbf{X} = \mathbf{U} \mathbf{L} \mathbf{V}^T \tag{14}$$

denote the singular value decomposition of $\mathbf{X}$ where $\mathbf{U}$ and $\mathbf{V}$ are $N \times r$ and $p \times r$ matrices, respectively, both with orthonormal columns, and $\mathbf{L}$ is the $r \times r$ diagonal matrix with singular values as its entries listed in decreasing order.

For notational convenience we denote the sample version of $C$ by $\widehat{C}$. Combining (7) and (14) we obtain

$$\widehat{C} = \left(\mathbf{X}^T \mathbf{X}\right)^{-1/2} \left(\mathbf{X}^T \mathbf{Y}_c\right) \left(\mathbf{Y}_c^T \mathbf{Y}_c\right)^{-1/2} = \mathbf{V} \mathbf{U}^T \mathbf{Y}_c \left(\mathbf{Y}_c^T \mathbf{Y}_c\right)^{-1/2}, \tag{15}$$

where $\mathbf{Y}_c$ is the centred $\mathbf{Y}$. Similarly, the sample version $\widehat{\mathbf{b}}$ of the vector $\mathbf{b}$ in (13) is calculated by

$$\widehat{\mathbf{b}} = \frac{1}{\widehat{\kappa}_1} \left(\mathbf{X}^T \mathbf{X}\right)^{-1/2} \widehat{C} \widehat{\mathbf{g}}_1 = \frac{1}{\widehat{\kappa}_1} \mathbf{V} \mathbf{L}^{-1} \mathbf{U}^T \mathbf{Y}_c \left(\mathbf{Y}_c^T \mathbf{Y}_c\right)^{-1/2} \widehat{\mathbf{g}}_1, \tag{16}$$

where $\widehat{C} \widehat{\mathbf{g}}_1 = \widehat{\kappa}_1 \widehat{\mathbf{h}}_1$ is the sample version of (11) with $\widehat{\kappa}_1$ and $\widehat{\mathbf{g}}_1$ the first eigenvalue and right eigenvector respectively. The expressions (15) and (16) do not depend on the relationship between $p$ and $N$, so can be equally applied to conventional data with $p < N$ as well as to HDLSS data.

### *3.3. Pursuit of Interesting Dimensions*

Since the 1970s and more particularly since Friedman and Tukey (1974), subspaces in data are regarded 'uninteresting' if they are random or unstructured, and conversely, a projection, or subspace is interesting if it is far from Gaussian. Following these ideas for a given data structure $\mathbf{X}$, which could denote the original data, ranked data or a subset of the variables of the original data, we aim to find the subspace of those variables of $\mathbf{X}$ which contain interesting non-Gaussian structure. The first principal component does not normally contain enough structure, since it is one-dimensional , while the whole data contain structure and randomness. The goal is thus to search for a low-dimensional subspace which contains the essential information in the data.

Two closely related methods, Projection Pursuit and Independent Component Analysis, find non-Gaussian projections in data. See Huber (1985), Jones and Sibson (1987), Friedman (1987) and Hyvärinen, *et al* (2001) for good accounts of how to find one or more projections.

More recently Scholz *et al* (2004) and Koch and Naito (2007) proposed methods for determining the number of these projections or the dimension of the subspace. Motivated by their sub-Gaussian metabolite data, the 'optimal' dimension in Scholz *et al* (2004) is that which results in the highest number of independent components with negative kurtosis. Koch and Naito (2007) use kurtosis and skewness for their dimension selection method which is generally applicable and has a theoretical foundation. For this reason we will apply their criterion in the pursuit of the best subspace dimension. We briefly review the relevant parts of their method and relate their theoretical results in Section 4.3.

Koch and Naito's starting point is the observation that the most interesting lower-dimensional subspace is that in which the largest deviation from the Gaussian can be obtained. An appropriate measure, which can be calculated directly from the data, is kurtosis. Let $\mathbf{x}_{p,j}$ with $j = 1, \ldots, N$ denote $p$-dimensional observations, here regarded as columns of $\mathbf{X}^T$. The sample kurtosis $\widehat{\beta}$ is defined by

$$\widehat{\beta}(\mathbf{X}) = \widehat{\beta}(\mathbf{x}_{p,1}, ..., \mathbf{x}_{p,N}) = \max_{\alpha \in \mathcal{U}^{p-1}} \widehat{B}(\alpha | \mathbf{x}_{p,1}, ..., \mathbf{x}_{p,N}), \qquad (17)$$

where

$$\widehat{B}(\alpha | \mathbf{x}_{p,1}, ..., \mathbf{x}_{p,N}) = \left| \frac{1}{N} \sum_{i=1}^{N} \left\{ \frac{\alpha^T(\mathbf{x}_{p,i} - \overline{\mathbf{x}}_p)}{\sqrt{\alpha^T S \alpha}} \right\}^4 - 3 \right|,$$

$\mathcal{U}^{p-1}$ is the unit sphere in $\mathbb{R}^p$, $S$ is the sample covariance of $\mathbf{X}$, and $\overline{\mathbf{x}}_p$ is the sample mean which is zero due to centring of $\mathbf{X}$. For $2 \le k \le p$, put $\tilde{\mathbf{Z}}^{(k)} = \mathbf{Z}^{(k)} \Lambda_k^{-1/2}$ with $\mathbf{Z}^{(k)}$ as in (4) and $\Lambda_k$ the covariance matrix of $\mathbf{Z}^{(k)}$. We calculate the sample kurtosis of the $\tilde{\mathbf{Z}}^{(k)}$, denoted by $\widehat{\beta}_k(\tilde{\mathbf{Z}}^{(k)})$. The most non-Gaussian dimension is the $H$ which satisfies

$$H = \operatorname{argmax}_{2 \le k \le p} \left\{ \sqrt{\frac{N}{4!}} \widehat{\beta}_k(\tilde{\mathbf{Z}}^{(k)}) - \tau_k \right\}, \qquad (18)$$

where $\tau_k$ is a certain constant used for bias-adjustment.

The choice of the most non-Gaussian dimension in Koch and Naito's method requires a sequence of subspaces of dimensions $1 < k \le p$. Principal component analysis (PCA) does provide such a sequence, but PCA is based purely on variance, and may therefore exclude variables that are important for regression prediction. For this reason, we cannot apply their criterion directly to the $k$-dimensional sphered PC data. Step 2 of our algorithm shows how the selection is achieved in a regression framework. Details relating to the choice of $\tau_k$ are given in Section 4.3.

### *3.4. A Regression Algorithm*

In this section we present our algorithm which consists of three steps. Step 1 achieves variable ranking which plays an important role in our algorithm and is one of the reasons why the phrase *supervised* can be used.

An effective dimension reduction method is applied in Step 2 and prediction as described in (5) is accomplished in the final Step 3.

*3.4.1. Step 1*

Let $[\mathbf{X}\ \mathbf{Y}]$ denote the predictor and response variables which satify (9) and (8). Let $\widehat{C}$ and $\widehat{\mathbf{b}}$ denote the ranking matrix and ranking vector obtained from $[\mathbf{X}\ \mathbf{Y}]$ as in (15) and (16). For $m = 2, \ldots, p$ let

$$|\widehat{b}_{j_1}| \geq |\widehat{b}_{j_2}| \geq \cdots \geq |\widehat{b}_{j_m}|$$

denote the $m$ largest entries (in absolute value) of $\widehat{\mathbf{b}}$, ranked in decreasing order, and define the $m$-dimensional ranked data

$$\mathbf{X}_m = [\mathbf{x}_{j_1} \cdots \mathbf{x}_{j_m}] = \begin{bmatrix} \mathbf{x}_{m,1}^T \\ \vdots \\ \mathbf{x}_{m,N}^T \end{bmatrix} \quad (19)$$

where the $i$-th variable is the variable corresponding to the $i$-th largest entry $\widehat{b}_{j_i}$ of $\widehat{\mathbf{b}}$. Our notation indicates that the column vectors $\mathbf{x}_{j_i}$ are feature vectors while the row vectors $\mathbf{x}_{m,k}^T$ correspond to the $N$ observations.

*3.4.2. Step 2*

For $m = 2, \ldots, \min(N, p)$ apply principal component analysis to $\mathbf{X}_m$. Sphere the principal components to obtain the $N \times m$ ranked and sphered PC data

$$\mathcal{S}_m = \mathbf{X}_m \boldsymbol{\Gamma} \boldsymbol{\Lambda}^{-1/2}, \quad (20)$$

where $N^{-1}\mathbf{X}_m^T\mathbf{X}_m = \boldsymbol{\Gamma}\boldsymbol{\Lambda}\boldsymbol{\Gamma}^T$ denotes the spectral decomposition with the eigenvalues in $\boldsymbol{\Lambda}$ arranged in decreasing order, $\boldsymbol{\Gamma}$ is the orthogonal matrix whose columns are eigenvectors belonging to the elements of $\boldsymbol{\Lambda}$.

For $m$ and the ranked and sphered PC data $\mathcal{S}_m$ calculate the sample kurtosis $\widehat{\beta}_k(\mathbf{X}_m\boldsymbol{\Gamma}_k\Lambda_k^{-1/2})$ for each $k \leq m$, with $\Lambda_k$ the appropriate covariance matrix, and determine the dimension $H = H(m)$ as in (18). Put

$$\widetilde{\mathbf{X}}_{m,H} = \mathbf{X}_m \boldsymbol{\Gamma}_H,$$

where $\boldsymbol{\Gamma}_H$ denotes the first $H$ columns of $\boldsymbol{\Gamma}$, and so $\widetilde{\mathbf{X}}_{m,H}$ is the matrix which consists of the first $H$ principal components of $\mathbf{X}_m$.

*3.4.3. Step 3*

For $m = 2, \ldots, \min(N, p)$ and a typical explanatory variable $\mathbf{z} \in \mathbb{R}^p$ predict the $q$-variate $\widehat{\mathbf{y}}$ as in (3) from the ranked data $\mathbf{X}_m$:

$$\widehat{\mathbf{y}}^T = \mathbf{z}_m^T \boldsymbol{\Gamma}_H \left( \widetilde{\mathbf{X}}_{m,H}^T \widetilde{\mathbf{X}}_{m,H} \right)^{-1} \widetilde{\mathbf{X}}_{m,H}^T \mathbf{Y} = \mathbf{z}_m^T \boldsymbol{\Gamma}_H \left( \boldsymbol{\Gamma}_H^T \mathbf{X}_m^T \mathbf{X}_m \boldsymbol{\Gamma}_H \right)^{-1} \boldsymbol{\Gamma}_H^T \mathbf{X}_m^T \mathbf{Y}, \quad (21)$$

where $\mathbf{z}_m$ is a subvector of $\mathbf{z}$ containing the first $m$ ranked variables as defined in (19) in Step 1.

## 4. Some Properties

### 4.1. Remarks on the Model and Algorithm

#### 4.1.1. Remarks on Step 1.

Our model is based on a set of features $\mathbf{X}_*$ which are described by $\mathbf{X}_m$ in the algorithm. This means obtaining $\mathbf{X}_m$ from $\mathbf{X}$ via $\widehat{C}$ is equivalent to finding the set of indices $I$ which characterise $\mathbf{X}_*$.

In addition to the ranking with $\widehat{\mathbf{b}}$ we also rank the data with the vector $\widehat{\mathbf{h}}_1$ which is the sample version of $\mathbf{h}_1$ in (11); see also (16). If a distinction between these two ranking schemes is required we put

$$\widehat{\mathbf{b}}_1 = \widehat{\mathbf{b}} \quad \text{and} \quad \widehat{\mathbf{b}}_2 = \frac{1}{\widehat{\kappa}_1}\widehat{C}\widehat{\mathbf{g}}_1 = \widehat{\mathbf{h}}_1.$$

Although $\widehat{\mathbf{b}}_1$ and $\widehat{\mathbf{b}}_2$ will lead to different submatrices $\mathbf{X}_m$, the rest of our algorithm proceeds in the same way for both ranking schemes. A comparison between the two ranking vectors shows that, apart from the scale factor $1/\widehat{\kappa}_1$, $\widehat{\mathbf{b}}_1$ contains sphering with $\left(\mathbf{X}^T\mathbf{X}\right)^{-1/2}$, therefore applies to raw or centred data, while $\widehat{\mathbf{b}}_2$ could be interpreted as applicable to data where sphering is not required or appropriate.

With very large data sets in particularly, it is not clear whether the sphering component is required, and we therefore propose to use both versions in applications.

Our ranking extends the ranking proposed by Bair *et al* (2006) which is essentially based on the correlation coefficient between the univariate response and each feature vector. We will return to these ranking schemes in Section 4.2.

#### 4.1.2. Remarks on Step 2.

The ranked feature set $\mathbf{X}_m$ is generally still too large to find the 'best' non-Gaussian predictors corresponding to $\mathbf{s}$ in (10). We use dimension reduction with PCA to pursue the best non-Gaussian subspace.

Using the notation of (20), we denote the eigenvalues of $N^{-1}\mathbf{X}_m^T\mathbf{X}_m$ by $\lambda_1 \geq \lambda_2 \geq \cdots \geq \lambda_m \geq 0$, and the eigenvectors by $\mathbf{\Gamma} = [\boldsymbol{\gamma}_1 \cdots \boldsymbol{\gamma}_m]$. It follows that

$$N^{-1}\mathbf{X}_m^T\mathbf{X}_m\boldsymbol{\gamma}_j = \lambda_j\boldsymbol{\gamma}_j \quad j = 1, ..., m, \tag{22}$$

and hence

$$N^{-1}\mathbf{X}_m^T\mathbf{X}_m = \sum_{j=1}^{m} \lambda_j \boldsymbol{\gamma}_j \boldsymbol{\gamma}_j^T = \sum_{j=1}^{H} \lambda_j \boldsymbol{\gamma}_j \boldsymbol{\gamma}_j^T + \sum_{j=H+1}^{m} \lambda_j \boldsymbol{\gamma}_j \boldsymbol{\gamma}_j^T.$$

In Step 2 we choose the cut-off '$H$' of (18) as in Koch and Naito (2007); so $H$ is the dimension which yields the most non-Gaussian projection. For this selection of $H$ we obtain $\widetilde{\mathbf{X}}_{m,H} = \mathbf{X}_m\mathbf{\Gamma}_H$. Further, by applying (18) to subspaces of the ranked rather than the original data, the most non-Gaussian dimension is determined from amongst the variables that are important for prediction.

*4.1.3. Remarks on Step 3.*

Instead of (1) we consider the multivariate regression model

$$\mathbf{Y} = \widetilde{\mathbf{X}}_{m,H}\mathbf{B}_r + \mathbf{E}, \tag{23}$$

where $\mathbf{B}_r$ is the $H \times q$ matrix of coefficients for subsets of the ranked data, and $\mathbf{E}$ denotes the errors. The estimate of $\mathbf{B}_r$ corresponding to (2) is

$$\widehat{\mathbf{B}}_r = \left(\widetilde{\mathbf{X}}_{m,H}^T \widetilde{\mathbf{X}}_{m,H}\right)^{-1} \widetilde{\mathbf{X}}_{m,H}^T \mathbf{Y}.$$

Hence for a datum $\mathbf{z}$, we extract the subvector $\mathbf{z}_m$ whose components correspond to the indices $\{j_1, ..., j_m\}$ obtained in Step 1. The prediction for $\mathbf{z}$ via the model (23) is obtained by $\widehat{\mathbf{y}} = \widehat{\mathbf{B}}_r^T \mathbf{\Gamma}_H^T \mathbf{z}_m$, which is nothing other than (21). Note that we can also express $\widehat{\mathbf{y}}$ in terms of the eigenvalues and eigenvectors of $N^{-1}\mathbf{X}_m^T \mathbf{X}_m$ as follows:

$$\widehat{\mathbf{y}} = \frac{1}{N}\sum_{j=1}^{H} \frac{1}{\lambda_j} \mathbf{Y}^T \mathbf{X}_m \boldsymbol{\gamma}_j \boldsymbol{\gamma}_j^T \mathbf{z}_m. \tag{24}$$

### 4.2. Properties of $\widehat{\mathbf{b}}$

For univariate responses, so $q = 1$, the regression model (1) reduces to

$$\mathbf{y} = \mathbf{X}\boldsymbol{\beta} + \boldsymbol{\varepsilon}, \tag{25}$$

where $\mathbf{y}$ is the $N \times 1$ response vector, $\boldsymbol{\beta}$ is the $p \times 1$ coefficient vector, and $\boldsymbol{\varepsilon}$ is the $N \times 1$ error vector. In this case, from (15),

$$\widehat{C} = \left(\mathbf{X}^T\mathbf{X}\right)^{-1/2} \mathbf{X}^T \mathbf{y}_c (\mathbf{y}_c^T \mathbf{y}_c)^{-1/2},$$

and hence we have

$$\widehat{\mathbf{h}}_1 = \frac{1}{\sqrt{\mathbf{y}_c^T \mathbf{X} \left(\mathbf{X}^T\mathbf{X}\right)^{-1} \mathbf{X}^T \mathbf{y}_c}} \left(\mathbf{X}^T\mathbf{X}\right)^{-1/2} \mathbf{X}^T \mathbf{y}_c,$$

where $\mathbf{y}_c$ is the centred response. Therefore it follows that

$$\widehat{\mathbf{b}} = \left(\mathbf{X}^T\mathbf{X}\right)^{-1/2} \widehat{\mathbf{h}}_1 \propto \left(\mathbf{X}^T\mathbf{X}\right)^{-1} \mathbf{X}^T \mathbf{y}_c \equiv \widehat{\boldsymbol{\beta}}. \tag{26}$$

It is well known that $\widehat{\boldsymbol{\beta}}$ is characterised as the vector $\boldsymbol{\beta}$ which maximises the correlation between $\mathbf{y}$ and $\mathbf{X}\boldsymbol{\beta}$ (see Section 10.2.1 of Mardia *et al*, 1979). Moreover any scalar multiple of $\widehat{\boldsymbol{\beta}}$ also maximises the correlation. Therefore using $\widehat{\mathbf{b}}$ is essentially equivalent to using $\widehat{\boldsymbol{\beta}}$ in our variable ranking since a scalar multiple does not affect the ranking. For $q = 1$ we actually use

$$\widehat{\mathbf{b}}_1 = \widehat{\boldsymbol{\beta}} \quad \text{and} \quad \widehat{\mathbf{b}}_2 = \left(\mathbf{X}^T\mathbf{X}\right)^{1/2} \widehat{\boldsymbol{\beta}}.$$

We compare ranking with $\widehat{\mathbf{b}}$ to that used in Bair *et al* (2006). Our method for choosing relevant predictors is based on the full canonical correlation matrix $\widehat{C}$ between the predictors and the responses, or the full LSE. In contrast, Bair *et al* (2006) use correlation coefficients or separate LSEs for the univariate response and each univariate feature with a model similar to (10). This corresponds to considering only the marginals, while ignoring interactions, and admits the interpretation of a single LSE under the simplified linear model

$$\mathbf{y} = \beta_j \mathbf{x}_j + \boldsymbol{\varepsilon} \ \text{ for } j = 1, \ldots, p,$$

where $\boldsymbol{\varepsilon}$ is an error vector with mean zero and variance $\sigma^2 I_N$. In other words, they apply simple regression with $\mathbf{x}_j$ as a single predictor, although their underlying setting is that of multiple regression (25). They use $s_j = \mathbf{x}_j^T \mathbf{y}/\|\mathbf{x}_j\|$, with $j = 1, \ldots, p$, for their ranking, which can be obtained as a standardised LSE of the $\beta_j$s. Bair *et al* implement their sorting in an analogous way to Step 1 of our algorithm, and their threshold, which corresponds to the choice of $m$ in our setting, is chosen by cross-validation.

Next we briefly examine properties of the $s_j$s. Under the usual manipulations conditional on $\mathbf{x}$, it follows that

$$E[s_j] = \frac{1}{\|\mathbf{x}_j\|} \left[ \|\mathbf{x}_j\|^2 \beta_j + \sum_{i \neq j} \mathbf{x}_j^T \mathbf{x}_i \beta_i \right], \quad var[s_j] = \sigma^2.$$

In particular, this calculation shows that $s_j$ is not unbiased for $\beta_j$. On the other hand, since $\widehat{\mathbf{b}}_1 = [\widehat{b}_{11} \cdots \widehat{b}_{1p}]^T = \widehat{\boldsymbol{\beta}}$,

$$E\left[\widehat{b}_{1j}\right] = \beta_j, \quad \text{and} \quad var\left[\widehat{b}_{1j}\right] = \sigma^2 d_{jj},$$

where $d_{k\ell}$ is the $(k, \ell)$-th component of $\left(\mathbf{X}^T\mathbf{X}\right)^{-1}$. Note that $\widehat{\mathbf{b}}_2$ is not unbiased, but the variance of each of its components is $\sigma^2$. For $p$ fixed, standard asymptotic theory suggests that $\left(\mathbf{X}^T\mathbf{X}\right)^{-1} = O_p(N^{-1})$ as $N$ grows, and thus $\widehat{b}_{1j}$ can estimate $\beta_j$ more accurately than the biased $s_j$ in the selection of relevant predictors. It is also intuitively obvious that variable selection should include the correlation structure from all predictors.

A further advantage of our ranking method over the selection of variables with $s$ is that the $\widehat{\mathbf{b}}$-ranking is applicable to multivariate response variables, while there is no obvious extension of $s$ to a truely multivariate setting.

Since $\widehat{\mathbf{b}}$ takes into account the whole correlation structure, computations are more complex and more involved than those resulting in the calculations of the $s_j$s. The latter can be carried out very efficiently for any number of features as the complexity only grows with the number of variables. Choosing between these two ranking schemes therefore represents a compromise between computational efficiency and exploitation of the correlation structure.

### *4.3. Details on the Dimension Selection*

For $2 \leq k \leq p$ and $\tilde{\mathbf{Z}}^{(k)} = \mathbf{Z}^{(k)}\Lambda_k^{-1/2}$ as in (18) of Section 3.3, $\widehat{\beta}_k(\tilde{\mathbf{Z}}^{(k)})$ increases with the dimension $k$. This means that direct use of $\widehat{\beta}_k(\tilde{\mathbf{Z}}^{(k)})$ is not useful for the selection of the best $k$.

Mimicking the derivation of the AIC and using some insight about null structure motivate us to consider the behaviour of

$$\sqrt{\frac{N}{4!}}\widehat{\beta}_k\left(\tilde{\mathbf{Z}}^{(k)}\right)$$

under the $k$-dimensional standard Gaussian structure (see Sections 3.1 and 3.2 in Koch and Naito, 2007). Under Gaussian assumptions it is shown that

$$\sqrt{\frac{N}{4!}}\widehat{\beta}_k\left(\tilde{\mathbf{Z}}^{(k)}\right) \to T_k \quad \text{in distribution as } N \to \infty,$$

and hence also that

$$E\left[\sqrt{\frac{N}{4!}}\widehat{\beta}_k\left(\tilde{\mathbf{Z}}^{(k)}\right)\right] \simeq E\left[T_k\right]$$

for large $N$, where $T_k$ is the maximum of a zero mean Gaussian random field on $\mathcal{U}^{k-1}$.

The quantity $E[T_k]$ cannot be calculated directly, but is estimated via the bounds below, here only given for kurtosis and adjusted to our scenario.

**Theorem** [Koch and Naito (2007)]
*For each $k \leq \min\{p, N\}$*

$$LB_k \leq E[T_k] \leq UB_k$$

where

$$\begin{aligned}
LB_k &= \sum_{e=0, e:even}^{k-1} \omega_{k-e} \Lambda_{k-e,\rho-k+e}(\tan^2 \theta), \\
UB_k &= LB_k + \sqrt{0.6}\, E[\chi_\rho]\left[1 - \Psi(\theta, k)\right].
\end{aligned}$$

Since the lower bound LB vanishes rapidly with dimension, Koch and Naito continue with upper bound UB. The interested reader is referred to their paper for details on the lower bound. In the notation of their theorem, $\theta = \cos^{-1}(\sqrt{0.6})$, $\Psi$ is a weighted sum of upper tail probabilities of the beta distribution, and $\chi_\rho$ denotes a $\chi$-distributed random variable with $\rho$ degrees of freedom and $\rho = \binom{k+3}{4}$. Good approximations for the values of $UB_k$ are given in Koch and Naito (2007), and are denoted by $\widehat{UB}_k$. Koch and Naito propose to use these tabulated values as the constant $\tau_k$ in (18). Returning to the notation in Section 3.4.2, Step 2, we

- define, for $k \leq m$, the bias adjusted version $\widehat{I}_k$ of $\widehat{\beta}_k$ by

$$\widehat{I}_k = \sqrt{\frac{N}{4!}} \widehat{\beta}_k(\mathbf{X}_m \mathbf{\Gamma}_k \Lambda_k^{-1/2}) - \widehat{UB}_k,$$

- and take

$$H(m) = \operatorname{argmax}_{2 \leq k \leq m} \widehat{I}_k,$$

as the practical dimension selector of the ranked $m$-dimensional data $\mathcal{S}_m$ consisting of $N$ samples.

The theorem and the practical dimension selector show that the dimension $H(m)$ is that for which the difference to the Gaussian is maximal.

### 4.4. A General View of Supervised Principal Components

In this section we clarify why our algorithm can be regarded as a generalisation of the 'Prediction by supervised principal components' proposed by Bair *et al* (2006). Like our method, theirs starts with $\mathbf{X}_m$, which they obtain by their simpler variable ranking described in Subsection 4.2. We express the singular value decomposition (14) of $\mathbf{X}_m$ in terms of the individual eigenvalues and eigenvectors and obtain

$$\mathbf{X}_m = [\boldsymbol{u}_1 \cdots \boldsymbol{u}_m] \, diag\left\{\sqrt{d_1}, ..., \sqrt{d_m}\right\} \begin{bmatrix} \boldsymbol{v}_1^T \\ \vdots \\ \boldsymbol{v}_m^T \end{bmatrix},$$

where, for $j = 1, ..., m$, the $d_j$s, $\boldsymbol{u}_j$s and $\boldsymbol{v}_j$s satisfy

$$\mathbf{X}_m^T \mathbf{X}_m \boldsymbol{v}_j = d_j \boldsymbol{v}_j, \quad \mathbf{X}_m \mathbf{X}_m^T \boldsymbol{u}_j = d_j \boldsymbol{u}_j, \quad \mathbf{X}_m \boldsymbol{v}_j = \sqrt{d_j} \boldsymbol{u}_j \qquad (27)$$

In equations (6) and (7) of their Section 2.1, Bair *et al* (2006) describe their prediction $\widehat{y}^{spc}$ in a multiple regression model such as (25). For a given datum $\mathbf{z}_m$, and using our notation, this is essentially

$$\begin{aligned}
\widehat{y}^{spc} &= \left(\boldsymbol{u}_1^T \mathbf{y}\right) \mathbf{z}_m^T \left(\frac{1}{\sqrt{d_1}} \boldsymbol{v}_1\right) \\
&= \mathbf{z}_m^T \left(\frac{1}{\sqrt{d_1}} \boldsymbol{v}_1^T \mathbf{X}_m^T \mathbf{y}\right) \left(\frac{1}{\sqrt{d_1}} \boldsymbol{v}_1\right) \\
&= \mathbf{z}_m^T \left(\frac{1}{d_1} \boldsymbol{v}_1 \boldsymbol{v}_1^T\right) \mathbf{X}_m^T \mathbf{y} \\
&= \frac{1}{N\lambda_1} \mathbf{y}^T \mathbf{X}_m \boldsymbol{\gamma}_1 \boldsymbol{\gamma}_1^T \mathbf{z}_m,
\end{aligned} \qquad (28)$$

where the last equality can be seen using (22) and (27). Comparing (24) with (28), we see that $\widehat{y}^{spc}$ corresponds to the case $H = 1$ in (24), and the prediction in (24) is $q$-dimensional, while $\widehat{y}^{spc}$ is one-dimensional. In this sense our prediction is a natural generalisation of the method discussed by Bair *et al* (2006) to multivariate responses, where $q \geq 2$.

## 5. Results

This section reports on the numerical performance of our algorithm and comparisons of our method with that of Bair *et al* (2006). For the simulated data and the real data sets we predict $\widehat{\mathbf{y}}$. To assess and compare the performances of the different methods we use the least squares error criterion

$$\text{LSE}(m) = \left\{ \frac{1}{N} \sum_{i=1}^{N} ||\widehat{\mathbf{y}}_i - \mathbf{y}_i||^2 \right\}^{1/2} \tag{29}$$

where $m$ is number of variables of the ranked submatrix $\mathbf{X}_m$.

In Figures 1-5 we will adhere to the following notation on methods:

1. **knb1-pcH**: ranking with $\widehat{\mathbf{b}}_1$ followed by PCR with $H = H(m)$ selected components;
2. **knb2-pcH**: ranking with $\widehat{\mathbf{b}}_2$ followed by PCR with $H = H(m)$ selected components;
3. **bhpt-pcH**: ranking as in Bair *et al* (2006) followed by PCR with $H = H(m)$ selected components;
4. **bhpt-pc1**: ranking as in Bair *et al* (2006) followed by PCR with the first component only;
5. **nr-pcH**: PCR with $H = H(m)$ selected components based on the original data without ranking.

Here 'pcH' refers to Steps 2 and 3 of our algorithm, and 'pc1' refers to the prediction step of Bair *et al* (2006). We have added two other prediction methods, bhpt-pcH and nr-pcH. The first, bhpt-pcH, is a mixture of our method and that of Bair *et al*; it uses ranking as described in Bair *et al* (2006), but then applies Steps 2 and 3 of our algorithm. The last, nr-pcH, is only used in Example 3.2.1, and will be described further there. Like bhpt-pcH, it is used to assess the advantages of our variable ranking.

We calculate the predicted values $\widehat{\mathbf{y}}$ and the error (29) for $m \geq 2$. In each case we indicate which of the five methods have been used. We show performance plots with $m$ on the $x$-axis, and LSE on the $y$-axis.

### *5.1. Simulated Data*

Based on model (10) we generate two sets of data. The first refers to a classical multiple regression framework, and the second models a high dimension low sample size (HDLSS) general multivariate regression setting.

For both models we use the 3-dimensional source vector $\mathbf{s} = [s_1 \; s_2 \; s_3]^T$ whose components are independently distributed as follows: $s_1 \sim$ uniform on $[0, 1]$, $s_2 \sim$ Exponential with mean 1, and $s_3 \sim N(0, 1)$, the standard Gaussian. So $H = 3$ in the model. For the terms $\boldsymbol{\delta}$ and $\boldsymbol{\eta}$ we use vectors with independent components distributed as $0.5 \times N(0, 1)$, in $q$ dimensions for $\boldsymbol{\delta}$ and in $m$ dimensions for $\boldsymbol{\eta}$. The values for $q$ and $m$ will be different in the two settings as will be the transformations $\mathbf{P}$ and $\mathbf{W}$.

**Example 5.1.1: Classical Data** with $q = 1, p = 13$ and $N = 172$. We use the matrices

$$\mathbf{P} = \begin{bmatrix} I_3 & & \\ & 2I_3 & \\ 3 & 3 & 3 \end{bmatrix} \qquad (30)$$

and

$$\mathbf{W}^T = [4 \ -3 \ -2],$$

where $I_d$ denotes the $d \times d$ identity matrix. The predictor is constructed by stacking $\mathbf{x}^T = [\mathbf{x}_*^T \ \mathbf{x}_L^T]$ where $\mathbf{x}_L$ is a 6-dimensional vector, and each of the 6 components is generated independently as $0.5 \times N(1.5, 1)$. Since the matrix $\mathbf{P}$ gives rise to $m = 7$, the predictor vector has 13 dimensions. We generate 172 predictors and responses, and combine them to form one data set $D = \{(y_1, \mathbf{x}_1), ..., (y_{172}, \mathbf{x}_{172})\}$.

In the simulations we use 100 such data sets $D_j$. For each data set $D_j$ we calculate the predicted values $\widehat{\mathbf{y}}$ and the error (29) for $m = 2, \ldots, 13$ and each of the four prediction methods knb1-pcH, knb2-pcH, bhpt-pcH and bhpt-pc1.

Figure 1 shows the performance of the four methods for a typical simulation.

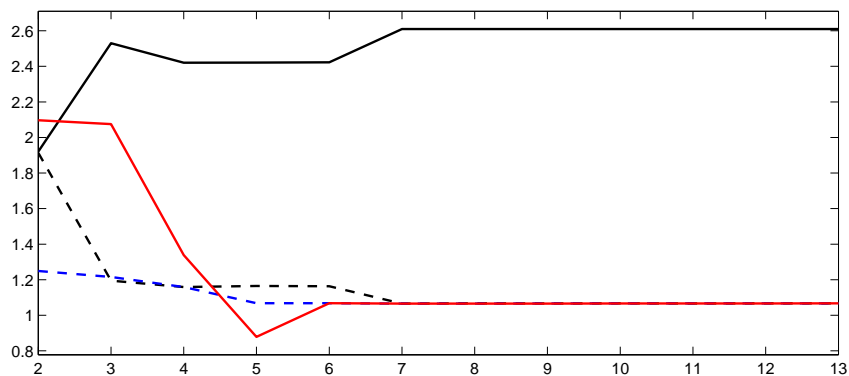

FIG 1. *LSE versus dimension for simulated data; knb1-pcH (red), knb2-pcH (blue), bhpt-pcH (black dashed) and bhpt-pc1 (black).*

The $x$-axis displays the dimension $m$ of $\mathbf{X}_m$ and the $y$-axis shows the LSE as in (29). All 100 simulations show that prediction with $H$ components outperforms prediction with the first component only, as seen in this figure. Using $H$ components leads to a strong decrease as dimenion increases and then the curves flatten out. In some cases, as for knb1-pcH in Figure 1, the curve has a minimum (here at 5) and then increases again. This behaviour is not uncommon and shows clearly that performance is not improved with more variables.

The next figure (Figure 2) examines in more detail the best final dimension $H$ for prediction. For each data set $D_j$ and for each method, say $M_i$, with $i =$ $1, \ldots, 4$, we find the dimension $m = m(j, i)$ and the associated final dimension $H = H(j, i)$ which minimises LSE. For bhpt-pc1 we equate $m$ and $H$, since only one dimension is selected in the process. Figure 2 shows the number of times out of 100 that a particular value $H = H(j, i)$ was determined as the dimension which resulted in the minimum LSE. The values for $H$ are shown on the $x$-axis and the $y$-axis shows the counts. So, for example, the final dimension $H = 2$ was found to produce the smallest error in more than 60 of the simulations with method knb1-pcH, while bhpt-pc1 resulted in smallest error less than 50 times for $m = H = 2$. The first three methods predominantly use 2 or 3 dimensions for best prediction. This agrees with our model in which $H = 3$ is used. Indeed, the simulations show that 2 out of these three variables are often sufficient for best prediction.

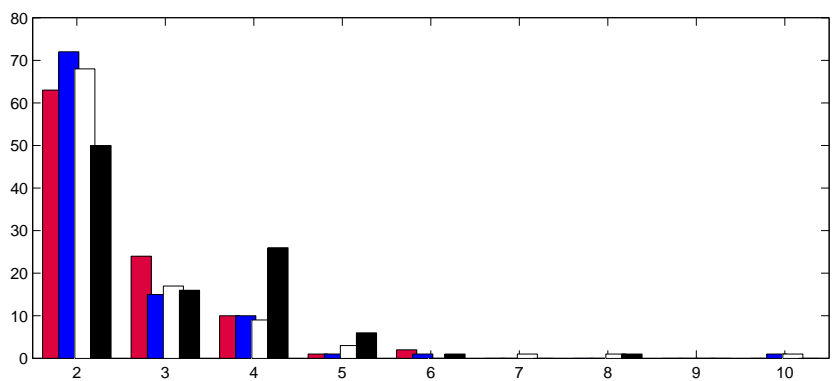

FIG 2. *Counts of best prediction versus final dimension $H$; knb1-pcH (red), knb2-pcH (blue), bhpt-pcH (white) and bhpt-pc1 (black) .*

We complement these two figures with Table 1 which shows, for each final dimension $H$, how many times out of 100 simulations a particular method resulted in the smallest LSE. We see that for $H = 2$ knb1-pcH had the smallest LSE in 37 simulations, while knb2-pcH only scored 18 times for $H = 2$. When there were ties for the smallest error, each method scored. For this reason the totals add up to more than 100. The last column of the table gives totals for each method. We see that knb1-pcH performed considerably better than either of the two other pcH methods. In none of the simulations did bhpt-pc1 perform best. For this reason bhpt-pc1 is not included in the Table 1.

Our results show that pcH methods, which are based on more than the first component, outperform bhpt-pc1. For the three pcH methods, ranking with $\widehat{\mathbf{b}}_1$ appears to produce much better results and at lower dimensions than the other two ranking methods which perform similarly in this classical setting.

**Example 5.1.2: HDLSS Data** with $q = 7, p = 172$ and $N = 52$. We use the model described in Example 5.1.1 to generate the data, but this time $N << p$. Instead of using the 6-dimensional vector $\mathbf{x}_L$ of the first example, $\mathbf{x}_L$ now has 165 dimensions. The matrix $\mathbf{P}$ remains the same, but $\mathbf{W}$ is replaced by

$$\mathbf{W}^T = \begin{bmatrix} 4 & -3 & -2 \\ & I_3 & \\ 1 & -2 & 0 \\ 0 & 1 & -2 \\ 1 & 0 & -2 \end{bmatrix}$$

to allow for multivariate responses.

|          | 2  | 3  | 4 | 5 | 6-8 | total |
|----------|----|----|---|---|-----|-------|
| knb1-pcH | 37 | 13 | 6 | 1 | 0   | **57** |
| knb2-pcH | 18 | 8  | 1 | 0 | 1   | **28** |
| bhpt-pcH | 20 | 9  | 3 | 2 | 3   | **37** |

TABLE 1
Counts of best performances over all methods by dimension $H$. Ties account for the total of more than 100. PCR never performed best.

We carried out 25 simulations and calculated the multivariate predictions and the error based on the rankings with $\widehat{\mathbf{b}}_1$ and $\widehat{\mathbf{b}}_2$. Since the ranking of Bair *et al* (2006) applies to univariate responses only, we can not use bhpt-pc1 or bhpt-pcH.

The sample size $N = 52$ limits the rank of $\mathbf{X}_m$, so we can only consider dimensions $m \leq 52$. The best dimension selector of Koch and Naito (2007) contains values up to 50 dimensions, and for this reason we restrict $\mathbf{X}_m$ to maximally 50 variables. Figure 3 shows the performance of 2 typical simulations. The $x$-axis shows the dimension $m$ against the LSE for knb1-pcH and knb2-pcH on the $y$-axis.

Unlike the previous example, for the HDLSS data the performance with $\widehat{\mathbf{b}}_2$ is either en par or better than that of $\widehat{\mathbf{b}}_1$. For this reason we have shown the results of two simulations in Figure 3. Most performance curves had a sharp drop (big improvement in performance) at a specific dimension $m$ rather than a gradual decrease in error. The final dimension $H$ corresponding to the smallest error was at most 8 in both methods and all simulations. Table 2 shows results analogous to those of Figure 2 for the HDLSS example but this time as percentages.

|          | 2  | 3  | 4  | 5  | 6  | 7 | 8 |
|----------|----|----|----|----|----|---|---|
| knb1-pcH | 32 | 24 | 12 | 12 | 12 | 4 | 4 |
| knb2-pcH | 36 | 36 | 4  | 12 | 12 | 0 | 0 |

TABLE 2
Percentage of best prediction by method for each dimension $H$.

Table 2 shows that two or three dimensions are mostly enough for good prediction. This outcome is similar to that of the previous example. Closer inspection of the table and Figure 3 reveals that ranking with $\widehat{\mathbf{b}}_2$ results more often in smaller error and at smaller dimenensions than ranking with $\widehat{\mathbf{b}}_1$. This is the opposite of what we observed in Example 5.1.1.

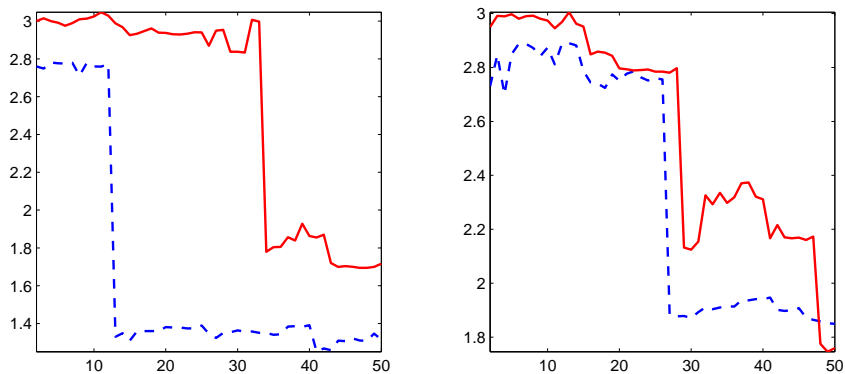

Fig 3. *LSE versus dimension; knb1-pcH (red) and knb2-pcH (blue).*

## *5.2. Applications to Real Data*

We consider two real data sets, one with multivariate responses and $N > p$, and our final example is a gene expression data set where the response variable consists of survival times. These data complement our simulations in that the first multivariate data set is classical while the microarray data has the large dimension of 24481 and only 78 samples.

**Example 5.2.1: Illicit drug market data** with $q = 7, p = 10$ and $N = 66$. The data contain monthly counts of events recorded by units of key health, law enforcement and drug treatment agencies in New South Wales, Australia. The data were collected over 66 months from January 1997 to June 2002 and consist of 17 different features which group into direct and indirect measures of the drug market. The data are described in Gilmour and Koch (2004).

The goal of the present analysis is to predict the 7 indirect measures of the drug market from the 10 direct measures.

The multivariate responses exclude use of bhpt-based ranking methods, instead we examine the effect of ranking, by comparing the two ranking schemes to non-ranked data. The crucial difference between our methods and nr-pcH is that $\mathbf{X}_m$ of Step 1 of our algorithm which contains the 'best' $m$ variables is replaced by the first $m$ variables of $\mathbf{X}$ without any sorting being applied, and so consist of the variables in the *natural* order in which they have been collected. Steps 2 and 3 of our algorithm remain the same in nr-pcH.

Figure 4 shows the performance of our two methods and nr-pcH with the dimension on the $x$-axis, and the LSE for the 7-dimensional responses on the $y$-axis. The methods based on ranking have a similar performance, and have generally smaller errors than nr-pcH, especially for smaller values of $m$. When all variables are used there is of course no difference in the performance, and the same value $H$, here $H = 8$ is selected in all three methods.

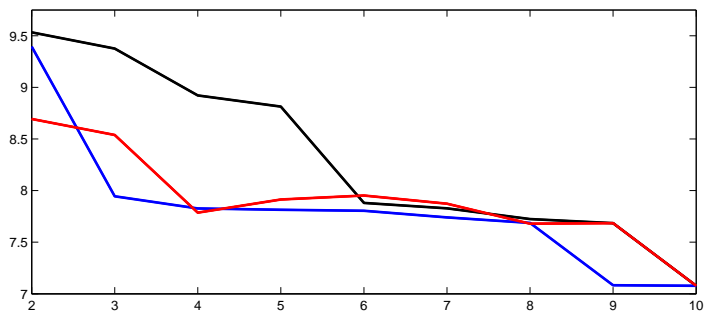

FIG 4. *LSE versus dimension for the illicit drug data with knb1-pcH (red),knb2-pcH (blue) and nr-pcH (black).*

Since the number of variables is small we show, in Table 3, the final dimension $H$ that has been chosen for each of the three methods. We note in Table 3 that the final dimension does not always increase monotonically, as for $m = 6$ and knb1-pcH. Higher values of $H$ for fixed $m$ tend to result in lower errors, but due to the iterative nature of `FastICA` which is used in Koch and Naito (2007) to select $H$, it is not always possible to find the mixing matrix $A$. In such cases we first increase the number of iterations, and if this does not produce the desired mixing matrix, we decrease $H$ by one.

| $m$ | 2 | 3 | 4 | 5 | 6 | 7 | 8 | 9 | 10 |
|---|---|---|---|---|---|---|---|---|---|
| knb1-pcH | 2 | 2 | 4 | 4 | 3 | 5 | 6 | 5 | 8 |
| knb2-pcH | 2 | 3 | 4 | 5 | 5 | 5 | 5 | 8 | 8 |
| nr-pcH | 2 | 2 | 4 | 5 | 4 | 4 | 5 | 5 | 8 |

TABLE 3
Dimensions $H$ for the illicit drug data with knb1-pcH, knb2-pcH and nr-pcH.

We propose to use both ranking schemes and to *let the data choose* which method is preferable. In summary the analysis of the illicit drug data demonstrates the effect of ranking in decreasing the error and improving performance.

**Example 5.2.2: Breast cancer survival data** with $q = 1, p = 24481$ and $N = 78$. Our final example uses microarray data of expression levels of approximately 25,000 genes of breast cancer patients. The first column of the data represents the time until metastasis (or the time until the patient left the study);

the second column is 1 if the tumor metastasised, and 0 if it did not metastasise within 5 years. There are 44 patients who survived beyond 5 years and 34 who did not.

The data are described in van 't Veer *et al* (2002) and reference to further analyses of these data are given there. Mostly these data have been used for classification. As our interest is prediction of a continuous variable, we work with the actual survival times – given in months – as the response variables. For prediction of the time variable, the distinction between the two classes of patients is not relevant in itself.

Although the time of 5 years to metastasis is of medical interest, the actual survival times and their accurate prediction will complement analysis based on classification and may lead to a better understanding of the genes that are associated with metastasis.

The response variable is univariate, and we will therefore analyse the data using the four methods knb1-pcH, knb2-pcH, bhpt-pcH and bhpt-pc1 as in Example 5.1.1. Our analysis focusses on two issues:

- the selection of genes from each method; and
- the LSE of each method.

**Selection of genes.** We work with a total of 24,481 gene expresseion levels and 78 patients. All our analyses involve PCA which automatically restricts the number of genes to at most 78. Subsequent analysis with ICA and the dimension selector of Koch and Naito (2007) yields dimension selection for up to 50-dimensional data. For this reason we focus on a final selection of the 50 most important genes in the analysis. Thus we want to select about 0.2% of the variables.

The sorting of variables used by Bair *et al* (2006) is independent of the total number of variables since it compares one variable at a time with the response and then orders all variables. The ranking with $\widehat{\mathbf{b}}_1$ and $\widehat{\mathbf{b}}_2$ is conceptually and computationally much more involved. Since we desire a relatively small proportion of genes, about 0.2% of a very large number, we include a preliminary ranking prior to Step 1 of our algorithm. The preliminary step is achieved by the following computations.

1. Divide the data into $L$ disjoint submatrices $\mathbf{X}^{[\ell]}$ of size $N \times s$.
2. For each $\ell = 1, \ldots, L$ determine the $\tau$ most relevant variables of $\mathbf{X}^{[\ell]}$ obtained from ranking with $\widehat{\mathbf{b}}$. This results in submatrices $\mathbf{X}^{[\ell]}_\tau$ of size $N \times \tau$.
3. Combine the pre-ranked submatrices and obtain

$$\mathbf{X}^{[\tau-rank]} = \left[ \mathbf{X}^{[1]}_\tau \ \mathbf{X}^{[2]}_\tau \ldots \mathbf{X}^{[L]}_\tau \right],$$

   a matrix of size $N \times L\tau$, and note that $L\tau < p$.
4. Use $\mathbf{X}^{[\tau-rank]}$ instead of the original data matrix $\mathbf{X}$ as the input matrix to Step 1.

The advantages of preliminary or $\tau$-ranking are two-fold. Genes which are not important for prediction can be eliminated early; and computations especially of the matrix $C$ in (15) are much more efficient without loss of relevant intercorrelation information.

Computations with the breast cancer data have shown that the choices of $L$, $s$, $\tau$ and the initial partitioning into submatrices of size $N \times s$ are not very crucial, but some preliminary calculations with a range of values for $L$, $s$ and $\tau$ should be carried out before one settles on specific values for the actual analysis. For fixed $L$ and $s$ we partitioned the data in different ways into the initial submatrices $\mathbf{X}^{[\ell]}$ and found that the overlap of genes in the resulting $\tau$-ranked sets was well over 70% and that of the finally selected 50 genes well above 60%. We found that $L = 5$, $s = 5000$ and $\tau = 200$ work well for the breast cancer data. The preliminary ranking and ranking of Step 1 of the algorithm always employ the *same* ranking, so both times either $\widehat{\mathbf{b}}_1$ or $\widehat{\mathbf{b}}_2$.

A comparison of the degree of communality between the different ranking methods is given in Table 4. This table shows the percentage of common genes between the variables selected for the three selection methods with $\widehat{\mathbf{b}}_1$, $\widehat{\mathbf{b}}_2$ and the ranking of Bair *et al* (2006). We have presented percentage overlaps for the 1000 variables we obtain in the preliminary ranking, followed by the 200 best in Step 1, and the subset of the final 50 genes which are used for prediction in Figure 5. In Bair *et al* (2006) the best 1000, 200 and 50 are obtained in their ranking as a one-step process. In the table, we compare a 1000 genes of one method with 1000 genes of the other methods and 50 with 50.

|      | knb1 |     |    | knb2 |     |    |
|------|------|-----|----|------|-----|----|
|      | 1000 | 200 | 50 | 1000 | 200 | 50 |
| knb2 | 58.6 | 55  | 46 | -    | -   | -  |
| bhpt | 17.7 | 9.5 | 8  | 33.8 | 15  | 14 |

TABLE 4
Percentage overlap of selected genes for best 1000, 200 and 50 variables for each ranking.

It is interesting to observe that $\widehat{\mathbf{b}}_1$ and $\widehat{\mathbf{b}}_2$ have 23 out of 50 genes in common for the final best genes. Bair *et al* (2006)'s ranking, on the other hand, appears to choose very different genes especially from those selected with $\widehat{\mathbf{b}}_1$. In the performance plots we will see how this choice of final genes affects the LSE.

**Prediction of survival times.** From the best 50 genes for each ranking methods we predict the survival time, and we calculate LSE for each of the four methods over the range of dimensions $m = 2, \ldots, 50$. The minimum LSE and corresponding $H$ for each method are given in Table 5

|        | knb1-pcH | knb2-pcH | bhpt-pcH | bhpt-pc1 |
|--------|----------|----------|----------|----------|
| $m$    | 50       | 40       | 35       | 20       |
| $H$    | 11       | 12       | 13       | -        |
| LSE($m$) | 13.84  | 12.23    | 17.80    | 20.56    |

TABLE 5
Smallest prediction error and best dimension for the breast cancer survival data.

The three pcH methods use very similar values for the best $H$, but their $m$-values differ. Here knb2-pcH has the smallest error, this is closely followed by knb1-pcH, while the best predictions of the other two methods are clearly worse.

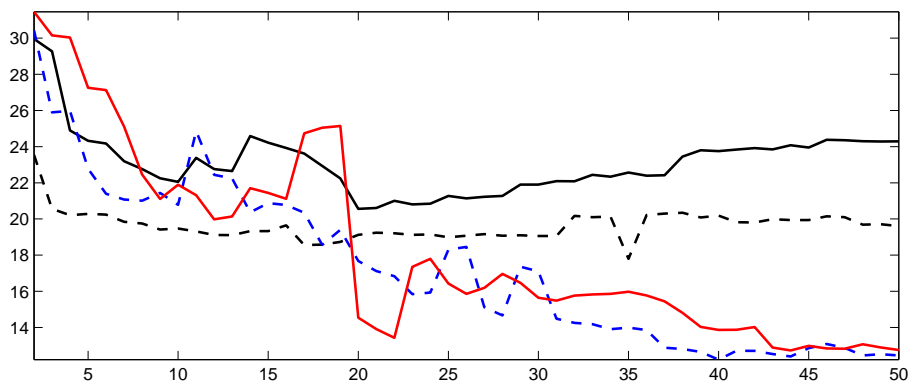

Fig 5. LSE versus dimension; knb1-pcH (red), knb2-pcH (blue), bhpt-pcH (black dashed) and bhpt-pc1 (black).

Our final performance results over the best 50 dimensions is shown in Figure 5 with the dimension $m$ on the $x$-axis and the LSE on the $y$-axis. We note that bhpt-pc1 is initially comparable to knb1-pcH and knb2-pcH until it reaches its best prediction at $m = 20$ from whence it increases again, while the two new competitors continue to decrease. The mixed method bhpt-pcH performs better initially than the other methods, but flattens out round $m = 20$. It does however have a clear minimum at $m = 35$ after which its LSE increases again. Overall knb1-pcH and knb2-pcH have lower LSEs. They show fairly similar behaviour; knb2-pcH has the smallest LSE and reaches its minimum earlier than knb1-pcH, and then increases slightly.

The results indicate that the 50 best genes were sufficient for prediction of the survival times, since all four methods had a minimum in this range of $m$-values and all apart from knb1-pcH showed a clear increase after having passed the minimum. The results also show that with the preliminary ranking prior to Step 1 of our algorithm our method can successfully be applied to HDLSS data with very large dimensions.

## 6. Discussion and Conclusion

This paper proposes a new method of supervised prediction in a regression setting which applies to multivariate responses, and to HDLSS problems as posed by gene expression data. The method intregrates variable ranking with a novel use of choosing the best number of predictors in PC regression, and extends current work by Bair *et al* (2006). The advantages of our comprehensive ranking are seen especially for large numbers of variables.

We demonstrate the performance of our method for multivariate responses in an HDLSS simulation and on real data. We test our approach against that of Bair *et al* (2006) in simulations and on microarray data of breast cancer survival times. The results convincingly show that our ranking combined with a careful selection of the number of components in PCR outperforms Bair *et al* (2006)'s method. The improved prediction comes however at a cost: The determination of the best dimensions, and the use of $H$ PCR components is computationally more expensive than PCR with the first component only.

Our results show the improved accuracy over existing prediction methods and open the way for further research into other variable ranking methods which could take into account data dependent information.

# References

[1] Bair, E., Hasie, T., Paul, D. and Tibshirani, R. (2006) Prediction by supervised principal components. *J. Amer. Statist. Assoc.*, **101**, 119-137.

[2] Friedman, J., and Tukey, J. W. T. (1974) A projection pursuit algorithm for exploratory data analysis. *IEEE Trans. Comput.*, **9**, 881-890.

[3] Friedman, J. H. (1987). Exploratory projection pursuit, *Journal of the American Statistical Association*, **82**, 249-266.

[4] Gilmour, S. and Koch, I. (2004) Understanding Australian Illicit Drug Markets: Unsupervised Learning with Independent Component Analysis. *Proceedings of the 2004 Intelligent Sensors, Sensor Networks & Information Processing Conference, ISSNIP*, Melbourne, Australia, 271-276.

[5] Gustafsson, M. G. (2005) Independent component analysis yields chemically interpretable latent variables in multivariate regression. *J. Chem. Inf. Model.*, **45**, 1244-1255.

[6] Hastie, T. J., Taylor, J. Tibshirani, R. and Walther, G. (2007) Forward stagewise regression and the monotone lasso. *Electronic Journal of Statistics*, **1**, 1-29.

[7] Huber, P. J. (1985) Projection pursuit (with discussion). *The Annals of Statistics.*, **13**, 435-475.

[8] Huang,D.-S. and Zheng,C.-H. (2006) Independent component analysis-based penalized discriminant method for tumor classification using gene expression data. *Bioinformatics*, **22**, 1855-1862.

[9] Hyvärinen, A. (1999) Fast and robust fixed-point algorithms for independent component analysis. *IEEE Trans. Neural Network*, **10**, 623-634.

[10] Hyvärinen, A., Karhunen, J. and Oja, E. (2001) *Independent Component Analysis*, Wiley, New York.

[11] Jones, M. C. and Sibson, R. (1987). What is Projection Pursuit? (with discussion) *The Journal of the Royal Statisitcal Society Series A*, **150**, 1-36.

[12] Koch, I. (2009). *Analysis of Multivariate and High-Dimensional Data*, to be published by Cambridge University Press, http://users.tpg.com.au/alunpope/inge/AMHDbook.pdf.
[13] Koch, I. and Naito, K. (2007) Dimension selection for feature selection and dimension reduction with principal and independent component analysis, *Neural Computation*, **19**, 513-545.
[14] Liebermeister, W. (2002) Linear modes of gene expression determined by independent component analysis. *Bioinformatics*, **18**, 51-60.
[15] Mardia, K. V., Kent, J. and Bibby, J. (1979) *Multivariate Analysis*, Academic Press.
[16] Meier, L. and Bühlmann, P. (2007) Smoothing $\ell_1$-penalized estimators for high-dimensional time-course data. *Electronic Journal of Statistics*, **1**, 597-615.
[17] Scholz, M., Gatzek, S., Sterling, A., Fiehn, O. and Selbig, J. (2004) Metabolite fingerprinting: detecting biological features by independent component analysis, *Bioinformatics*, **20**, 2447-2454.
[18] Schimizu, A., Hoyer, P. O., Hyvärinen, A. and Kerminen, A. (2006) A Linear Non-Gaussian Acyclic Model for Causal Discovery, *Journal of Machine Learning Research*, **7**, 2003-2030.
[19] van 't Veer, L. J., Dai, H., van de Vijver, M. J., He, Y. D., Hart, A. A.. M, Mao, M., Peterse, H. L., van der Kooy, K. Marton, M. J., Witteveen, A. T., Schreiber, G. J., Kerkhoven, R. M., Roberts, C., Linsley, P. S., Bernards, R. and Friend, S. H. (2002) Gene expression profiling predicts clinical outcome of breast cancer, *Nature* **415**, 530 - 536.